\documentclass[journal=nalefd,manuscript=letter,layout=traditional]{achemso}
\setkeys{acs}{etalmode=truncate,maxauthors=10}

\usepackage[version=3]{mhchem} 
\usepackage[normalem]{ulem}
\usepackage{color}
\usepackage[hidelinks]{hyperref}

\author{\"Onder G\"ul}\email{gulonder@gmail.com}
\affiliation{QuTech, Delft University of Technology, 2600 GA Delft, The Netherlands}
\alsoaffiliation{Kavli Institute of Nanoscience, Delft University of Technology, 2600 GA Delft, The Netherlands}
\author{Hao Zhang}
\affiliation{QuTech, Delft University of Technology, 2600 GA Delft, The Netherlands}
\alsoaffiliation{Kavli Institute of Nanoscience, Delft University of Technology, 2600 GA Delft, The Netherlands}
\author{Folkert K. de Vries}
\affiliation{QuTech, Delft University of Technology, 2600 GA Delft, The Netherlands}
\alsoaffiliation{Kavli Institute of Nanoscience, Delft University of Technology, 2600 GA Delft, The Netherlands}
\author{Jasper van Veen}
\affiliation{QuTech, Delft University of Technology, 2600 GA Delft, The Netherlands}
\alsoaffiliation{Kavli Institute of Nanoscience, Delft University of Technology, 2600 GA Delft, The Netherlands}
\author{Kun Zuo}
\affiliation{QuTech, Delft University of Technology, 2600 GA Delft, The Netherlands}
\alsoaffiliation{Kavli Institute of Nanoscience, Delft University of Technology, 2600 GA Delft, The Netherlands}
\author{Vincent Mourik}
\affiliation{QuTech, Delft University of Technology, 2600 GA Delft, The Netherlands}
\alsoaffiliation{Kavli Institute of Nanoscience, Delft University of Technology, 2600 GA Delft, The Netherlands}
\author{Sonia Conesa-Boj}
\affiliation{QuTech, Delft University of Technology, 2600 GA Delft, The Netherlands}
\alsoaffiliation{Kavli Institute of Nanoscience, Delft University of Technology, 2600 GA Delft, The Netherlands}
\author{Micha\l{} P. Nowak}
\affiliation{QuTech, Delft University of Technology, 2600 GA Delft, The Netherlands}
\alsoaffiliation{Kavli Institute of Nanoscience, Delft University of Technology, 2600 GA Delft, The Netherlands}
\alsoaffiliation{Faculty of Physics and Applied Computer Science, AGH University of Science and Technology, al. A. Mickiewicza 30, 30-059 Krak\'ow, Poland}
\author{David J. van Woerkom}
\affiliation{QuTech, Delft University of Technology, 2600 GA Delft, The Netherlands}
\alsoaffiliation{Kavli Institute of Nanoscience, Delft University of Technology, 2600 GA Delft, The Netherlands}
\author{Marina Quintero-P\'erez}
\affiliation{QuTech, Delft University of Technology, 2600 GA Delft, The Netherlands}
\alsoaffiliation{Netherlands Organisation for Applied Scientific Research (TNO), 2600 AD Delft, The Netherlands}
\author{Maja C. Cassidy}
\affiliation{QuTech, Delft University of Technology, 2600 GA Delft, The Netherlands}
\alsoaffiliation{Kavli Institute of Nanoscience, Delft University of Technology, 2600 GA Delft, The Netherlands}
\author{Attila Geresdi}
\affiliation{QuTech, Delft University of Technology, 2600 GA Delft, The Netherlands}
\alsoaffiliation{Kavli Institute of Nanoscience, Delft University of Technology, 2600 GA Delft, The Netherlands}
\author{Sebastian Koelling}
\affiliation{Department of Applied Physics, Eindhoven University of Technology, 5600 MB Eindhoven, The Netherlands}
\author{Diana Car}
\affiliation{QuTech, Delft University of Technology, 2600 GA Delft, The Netherlands}
\alsoaffiliation{Kavli Institute of Nanoscience, Delft University of Technology, 2600 GA Delft, The Netherlands}
\alsoaffiliation{Department of Applied Physics, Eindhoven University of Technology, 5600 MB Eindhoven, The Netherlands}
\author{S\'ebastien R. Plissard}
\affiliation{Kavli Institute of Nanoscience, Delft University of Technology, 2600 GA Delft, The Netherlands}
\alsoaffiliation{Department of Applied Physics, Eindhoven University of Technology, 5600 MB Eindhoven, The Netherlands}
\alsoaffiliation{CNRS-Laboratoire d'Analyse et d'Architecture des Syst\`emes (LAAS), Universit\'e de Toulouse, 7 avenue du colonel Roche, F-31400 Toulouse, France}
\author{Erik P.A.M. Bakkers}
\affiliation{QuTech, Delft University of Technology, 2600 GA Delft, The Netherlands}
\alsoaffiliation{Kavli Institute of Nanoscience, Delft University of Technology, 2600 GA Delft, The Netherlands}
\alsoaffiliation{Department of Applied Physics, Eindhoven University of Technology, 5600 MB Eindhoven, The Netherlands}
\author{Leo P. Kouwenhoven}\email{L.P.Kouwenhoven@TUDelft.nl}
\affiliation{QuTech, Delft University of Technology, 2600 GA Delft, The Netherlands}
\alsoaffiliation{Kavli Institute of Nanoscience, Delft University of Technology, 2600 GA Delft, The Netherlands}
\alsoaffiliation{Microsoft Station Q Delft, 2600 GA Delft, The Netherlands}

\title{Hard superconducting gap in InSb nanowires}

\begin{document}


\clearpage
\newpage
\begin{abstract}
Topological superconductivity is a state of matter that can host Majorana modes, the building blocks of a topological quantum computer. Many experimental platforms predicted to show such a topological state rely on proximity-induced superconductivity. However, accessing the topological properties requires an induced hard superconducting gap, which is challenging to achieve for most material systems. We have systematically studied how the interface between an InSb semiconductor nanowire and a NbTiN superconductor affects the induced superconducting properties. Step by step, we improve the homogeneity of the interface while ensuring a barrier-free electrical contact to the superconductor, and obtain a hard gap in the InSb nanowire. The magnetic field stability of NbTiN allows the InSb nanowire to maintain a hard gap and a supercurrent in the presence of magnetic fields ($\sim 0.5$ Tesla), a requirement for topological superconductivity in one-dimensional systems. Our study provides a guideline to induce superconductivity in various experimental platforms such as semiconductor nanowires, two dimensional electron gases and topological insulators, and holds relevance for topological superconductivity and quantum computation.
\end{abstract}

\vspace{0.5cm}
A topological superconductor can host non-Abelian excitations, the so-called Majorana modes forming the basis of topological quantum computation\cite{Read2000PRB,Kitaev2001,Fu2008PRL,Oreg2010PRL,Lutchyn2010PRL,Alicea2011NatPhys}. Both the non-Abelian property and the topological protection of Majoranas crucially rely on the energy gap provided by the superconducting pairing of electrons that separates the ground state from the higher energy excitations. For most material systems that can support such a topological state, pairing is artificially induced by proximity, where the host material is coupled to a superconductor in a hybrid device geometry\cite{Mourik2012Science,Das,Rokhinson,Deng2012,Churchill2013PRB,Finck2013PRL,Albrecht2016Nature,Deng2016,Zhang2016ArXiv,Chen2016ArXiv,Hart,Pribiag,Wang2012Science,Xu2014PRL,Xu2015PRL,Williams2012PRL,Wiedenmann2016,Bocquillon2016NatureNano,Wan2015NatCom,Pientka2016ArXiv,Hell2016ArXiv}. Accessing the topological properties in hybrid devices requires a negligible density of states within the induced superconducting gap, \textit{i.e.}, an induced hard gap, which can be attained by a homogeneous and barrier-free interface to the superconductor\cite{Takei2013,Chang2015NN,Krogstrup,Kjaergaard,Shabani2016}. However, achieving such interfaces remains an outstanding challenge for many material systems, constituting a major bottleneck for topological superconductivity. Here we engineer a high-quality interface between semiconducting InSb nanowires and superconducting NbTiN resulting in an induced hard gap in the nanowire by improving the homogeneity of the hybrid interface while ensuring a barrier-free electrical contact to the superconductor. Our transport studies and materials characterization demonstrate that surface cleaning dictates the structural and electronic properties of the InSb nanowires, and determines the induced superconductivity together with the wetting of the superconductor on the nanowire surface. We show that both the induced gap and the supercurrent in the nanowire withstands magnetic fields ($\sim 0.5$ Tesla), a requirement for topological superconductivity in one-dimensional systems.

InSb nanowires have emerged as a promising platform for topological superconductivity\cite{Mourik2012Science,Deng2012,Churchill2013PRB,Zhang2016ArXiv,Chen2016ArXiv} owing to a large spin-orbit coupling\cite{vanWeperenPRB,Kammhuber2017ArXiv}, a large g factor\cite{vanWeperenNL,Kammhuber}, and a high mobility\cite{Gul,Kammhuber,Li2016SciRep,Gill2016APL}. These ingredients, together with a high-quality interface to a magnetic field resilient s-wave superconductor, are necessary to maintain a finite topological gap in one dimension\cite{Oreg2010PRL,Lutchyn2010PRL,Potter2011PRB184520,Sau2012PRB064512}. The interface quality can be inferred using tunneling spectroscopy which resolves the induced superconducting gap for a tunnel barrier away from the interface. To date, tunneling spectroscopy studies on proximitized InSb nanowires have reported a significant density of states within the superconducting gap, a so-called soft gap, suggesting an inhomogeneous interface\cite{Mourik2012Science,Deng2012,Churchill2013PRB,Chen2016ArXiv}. These subgap states destroy the topological protection by allowing excitations with arbitrarily small energy. Soft gaps have been observed also in other hybrid systems for cases where tunneling spectroscopy is applicable\cite{Das,Finck2013PRL,Irie2016PRB155305,Su2016ArXiv}. For other cases, interface inhomogeneity is indirectly inferred from a decreased excess current or supercurrent due to a deviation from Andreev transport\cite{BTK}, a common observation in hybrid systems\cite{Hart,Pribiag,Bocquillon2016NatureNano}. A hard gap has recently been realized in epitaxial InAs-Al materials\cite{Chang2015NN,Krogstrup,Kjaergaard,Shabani2016}, and in Bi$_2$Se$_3$\cite{Wang2012Science} and Bi$_2$Te$_3$\cite{Xu2014PRL,Xu2015PRL} epitaxially grown on NbSe$_2$, where the interface inhomogeniety can be minimized. However these studies do not provide further insight into the soft gap problem in material systems for which either epitaxy remains a challenge or when a high structural quality does not guarantee a barrier-free interface (\textit{e.g.} due to carrier depletion). Here we tackle the soft gap problem in InSb nanowire devices by focusing on the constituents of a hybrid device realization which are crucial for the interface.

In general, realizing a hybrid device begins with surface preparation of the host material followed by the deposition of a superconductor. In host materials with low surface electron density or a small number of electronic subbands such as semiconductor nanowires, the correct surface preparation is of paramount importance to ensure a barrier-free coupling to the superconductor. Here we also adopt this procedure for our nanowires\cite{Car2014AdvMat} whose native surface oxide forms an insulating layer that has to be removed. We describe the details of the nanowire growth, fabrication, and measurement setup in the Supporting Information.

\begin{figure}
	\vspace{-30pt}
	\includegraphics[width=0.85\columnwidth]{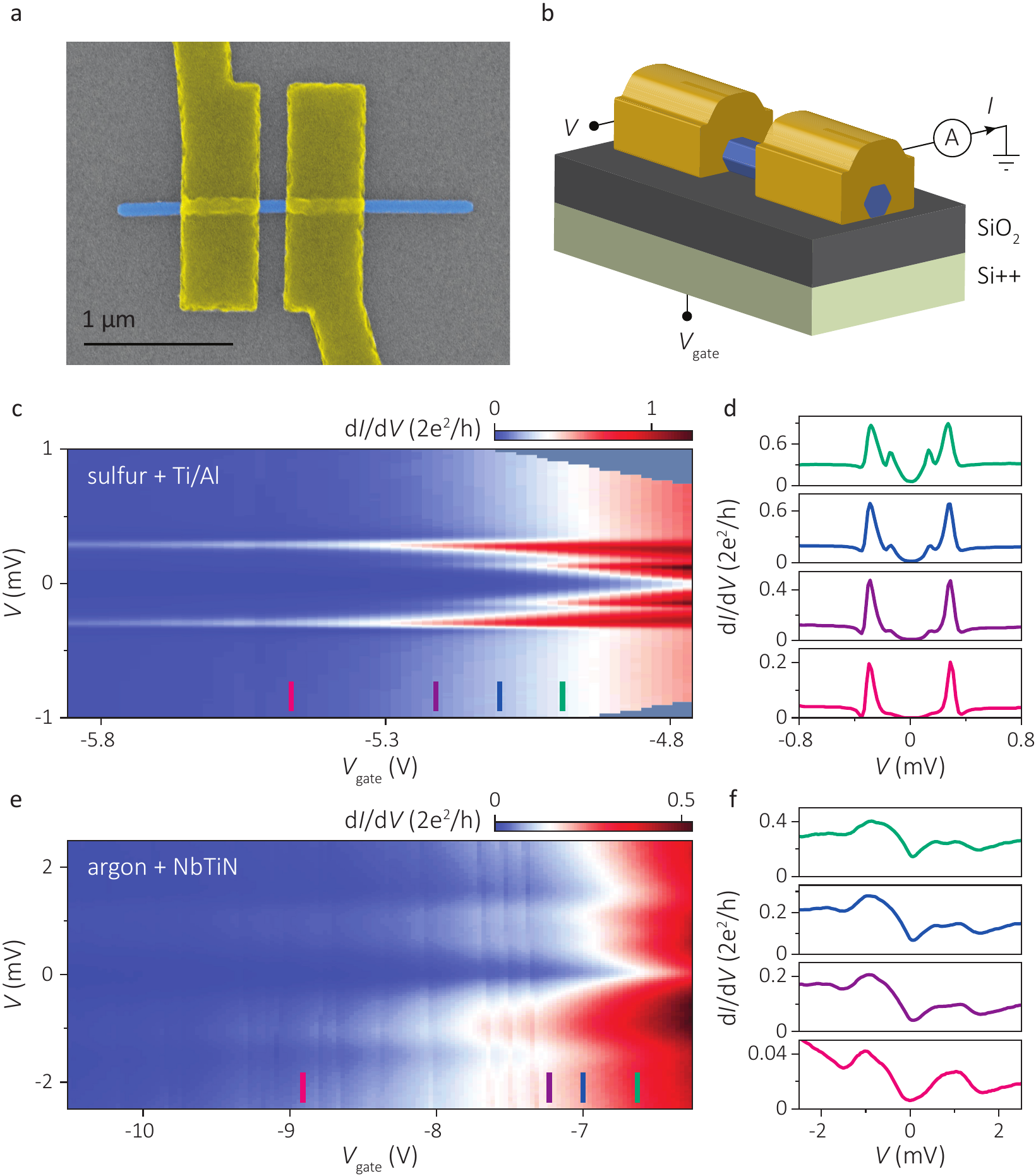}
	\caption{\footnotesize{\textbf{InSb nanowire hybrid device and induced superconducting gaps for different device realizations.} \textbf{(a)} Top-view false-color electron micrograph of a typical device consisting of an InSb nanowire (blue) with a diameter $\sim 80$ nm coupled to two superconducting electrodes (yellow) with $\sim 150$ nm separation. \textbf{(b)} Schematic of the devices and the measurement setup with bias voltage $V$, monitored current $I$, and the voltage $V_\mathrm{gate}$ applied on back gate (Si++ substrate) that is separated from the device by a 285 nm thick SiO$_2$ dielectric. \textbf{(c), (d)} Spectroscopy of a device realized using sulfur cleaning followed by evaporation of superconducting Ti/Al (5/130 nm) electrodes. $T = 250$ mK. Differential conductance d$I$/d$V$ is plotted as a function of bias voltage $V$ for varying gate voltages $V_\mathrm{gate}$. d$I$/d$V$ traces in (d) are vertical line cuts from (c) at gate voltages marked with colored bars. d$I$/d$V$ is symmetric around zero bias with two conductance peaks at $V \sim \pm 0.3$ mV seen for all gate voltages that result from the coherence peaks in the superconducting density of states at the edge of the induced gap $\Delta$. For our device geometry with two superconducting electrodes $2\Delta \sim 0.3$ meV. For sufficiently low $V_\mathrm{gate}$, where d$I$/d$V$ $\ll 2e^2/h$ at above-gap bias ($V > 2\Delta$), tunnelling is weak, which suppresses the Andreev reflection probability revealing a hard induced gap. Larger gate voltages decrease the tunnel barrier height where increased Andreev reflection probability results in finite subgap conductance. \textbf{(e), (f)} Spectroscopy of a device realized using argon cleaning followed by sputtering of superconducting NbTiN (90 nm) electrodes. $T = 250$ mK. We find $2\Delta \sim 1$ meV, much larger than that of the Al-based InSb hybrid device shown above. d$I$/d$V$ traces in (f) show an above-gap conductance comparable to those in (d). The induced gap is soft with a nonvanishing subgap conductance even for the weak tunnelling regime at low $V_\mathrm{gate}$, indicating a deviation from Andreev transport.}}
\end{figure}

Figure 1a and b show a completed device with two lithographically defined superconducting electrodes having a small separation ($\sim 150$ nm) on an InSb nanowire. A degenerately doped silicon substrate acts as a global back gate, tuning the carrier density in the wire. The small electrode separation allows us to electrostatically define a tunnel barrier in the wire section between the electrodes by applying negative gate voltages. Figure 1c and e show the induced gaps measured by tunneling spectroscopy for two common realizations of an InSb nanowire hybrid device. For the device in Figure 1c, a sulfur-based solution\cite{Suyatin2007} is used to clean the wire surface followed by evaporation of Ti/Al with Ti the wetting layer, whereas Figure 1e is from a device for which the wire surface is \textit{in-situ} cleaned using an argon plasma followed by sputtering of NbTiN. Figure 1d shows the conductance traces of the sulfur-Ti/Al device indicating a hard induced gap $2\Delta \sim 0.3$ meV for low gate voltages when decreased transmission suppresses Andreev reflection. In contrast, Figure 1f demonstrates that the argon-NbTiN device shows a soft induced gap even for the lowest gate voltages, but with a gap $2\Delta \sim 1$ meV inherited from NbTiN, a superconductor with a large gap and high critical field. Both device realizations present a challenge towards topological protection. In the first case, the magnetic field ($\sim 0.5$ T) required to drive the wire into the topological state destroys the superconductivity of Al (Figure S1). Al can withstand such fields when it is very thin ($< 10$ nm) in the field plane, however, such thin Al films contacting a nanowire have so far only been achieved by epitaxy\cite{Krogstrup,Albrecht2016Nature,Deng2016}. In the NbTiN device prepared with argon cleaning, the subgap states render the topological properties experimentally inaccessible.

\begin{figure}
	\includegraphics[width=1\columnwidth]{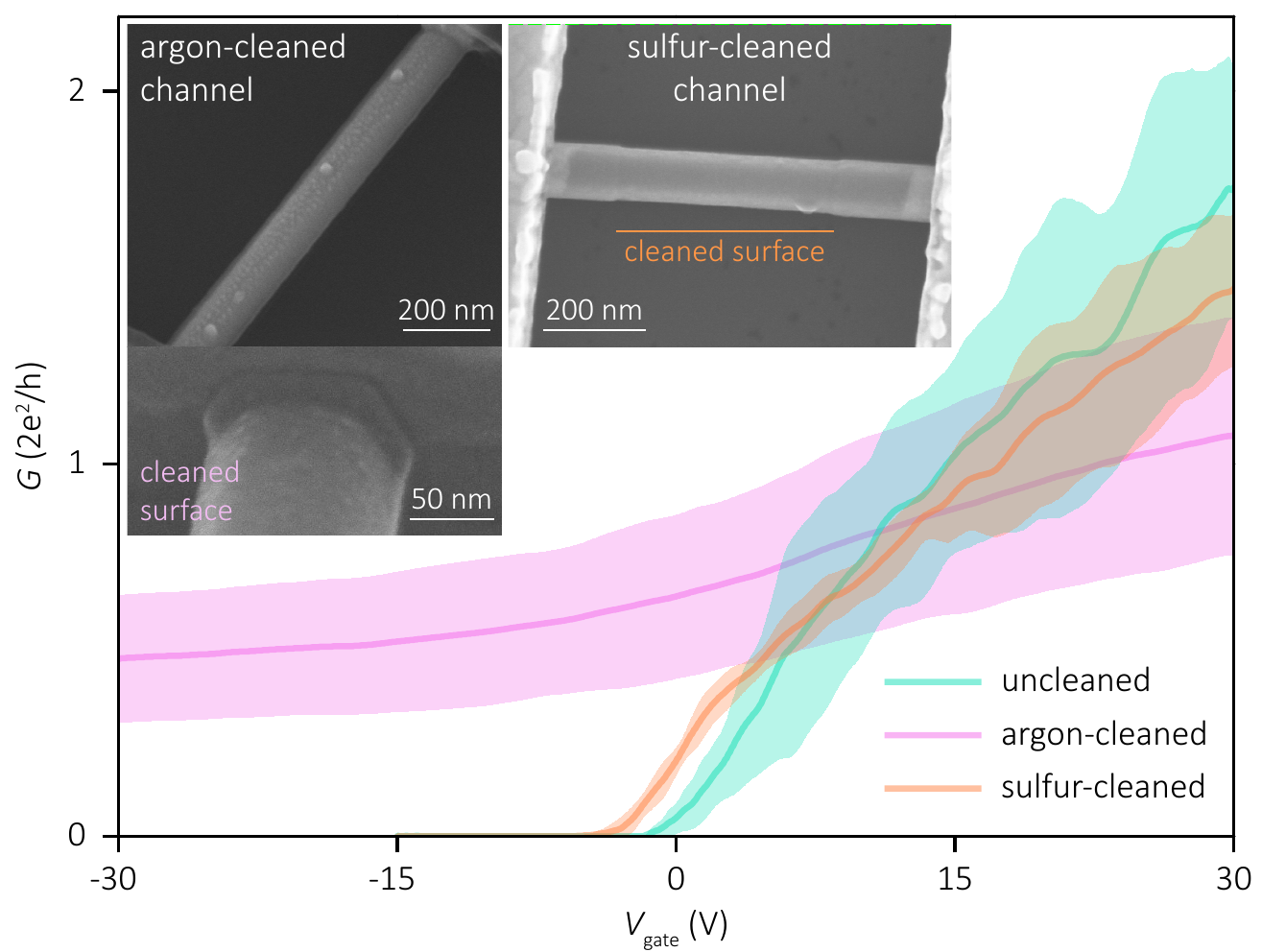}
	\caption{\footnotesize{\textbf{Effects of different surface cleaning on transport properties.} Gate voltage dependent conductance $G$ of InSb nanowire devices with $\sim 1$ $\mu$m electrode separation (channel length) for argon-cleaned (pink), sulfur-cleaned (orange), and uncleaned pristine (cyan) channels. $T = 4$ K. Traces represent ensemble-averaged conductance over 6 (argon-cleaned), 3 (sulfur-cleaned), and 2 (uncleaned) different devices measured at bias voltage $V = 10$ mV, with the shades indicating the standard deviation (see the Supporting Information for the details of averaging). Argon-cleaned channels do not pinch off, a deviation from a semiconducting gate response, and show a low transconductance $\propto$ d$G$/d$V_\mathrm{gate}$ indicating a low mobility. In contrast, sulfur-cleaned channels show a gate response similar to the pristine channel but with a shift of the threshold voltage towards negative values. Insets show high-resolution electron micrographs of argon- and sulfur-cleaned channels. Argon cleaning typically rounds the otherwise hexagonal cross section of the InSb nanowire (bottom image) and leaves a rough surface (top image). A sulfur cleaning yielding comparable contact resistances etches the InSb nanowire much less and leaves behind a smoother surface.}}
	\label{fig2}
\end{figure}

We now turn our attention to the surface of InSb nanowires prior to superconductor deposition. To determine the effects of surface cleaning on transport, we characterized long-channel nanowire devices with $\sim 1$ $\mu$m electrode separation where the channel surface is cleaned using different methods, along with control devices with pristine channels (details in the Supporting Information). Figure 2 shows the measured conductance through the nanowire as a function of gate voltage, with the traces representing an average over different devices and the shades indicating the standard deviation. We find that the argon-cleaned channel behaves strikingly different than sulfur-cleaned and pristine channels. First, the argon-cleaned channel does not pinch off, showing a finite conductance even for lowest gate voltages, indicating a deviation from a semiconducting gate response. Second, it shows a lower transconductance $\propto$ d$G$/d$V_\mathrm{gate}$ compared to sulfur-cleaned and pristine channels indicating a low mobility. These observations are consistent with the formation of metallic In islands on the InSb surface after argon cleaning\cite{Bouslama1995}. In contrast, the sulfur-cleaned channel shows a gate response similar to the pristine channel apart from a shift of the threshold voltage towards negative values. This behaviour indicates a surface electron accumulation expected for III-V semiconductors treated with sulfur-based solutions\cite{King2008,Petrovykh2003,Ho2009}. A close inspection of the cleaned channels reveals clear differences in nanowire surface morphology after argon and sulfur cleaning (Figure 2 inset). While argon cleaning created a roughness easily discernible under high-resolution electron microscope for different plasma parameters, we find that sulfur cleaning, which removes $\sim 5$ nm of the wire, leaves a smoother InSb surface. TEM studies on the cleaned wire surface confirm this observation (Figure S2). Comparable contact resistances between argon and sulfur cleaning were achievable (\textit{e.g.} in Figure 1e and f) when the argon plasma significantly etches the nanowire surface ($> 15$ nm), while different plasma parameters resulting in less etching gave consistently higher contact resistances. This indicates that a complete removal of the native oxide ($\sim 3$ nm) does not guarantee a barrier-free interface to the superconductor for InSb nanowires, which could be related to the surface depletion of InSb previously reported for a (110) surface\cite{Gobeli1965PR245}, the orientation of our nanowire facets. In the rest of the Letter we use sulfur cleaning to remove the native oxide on the nanowire surface prior to superconductor deposition.

\begin{figure}
	\includegraphics[width=1\columnwidth]{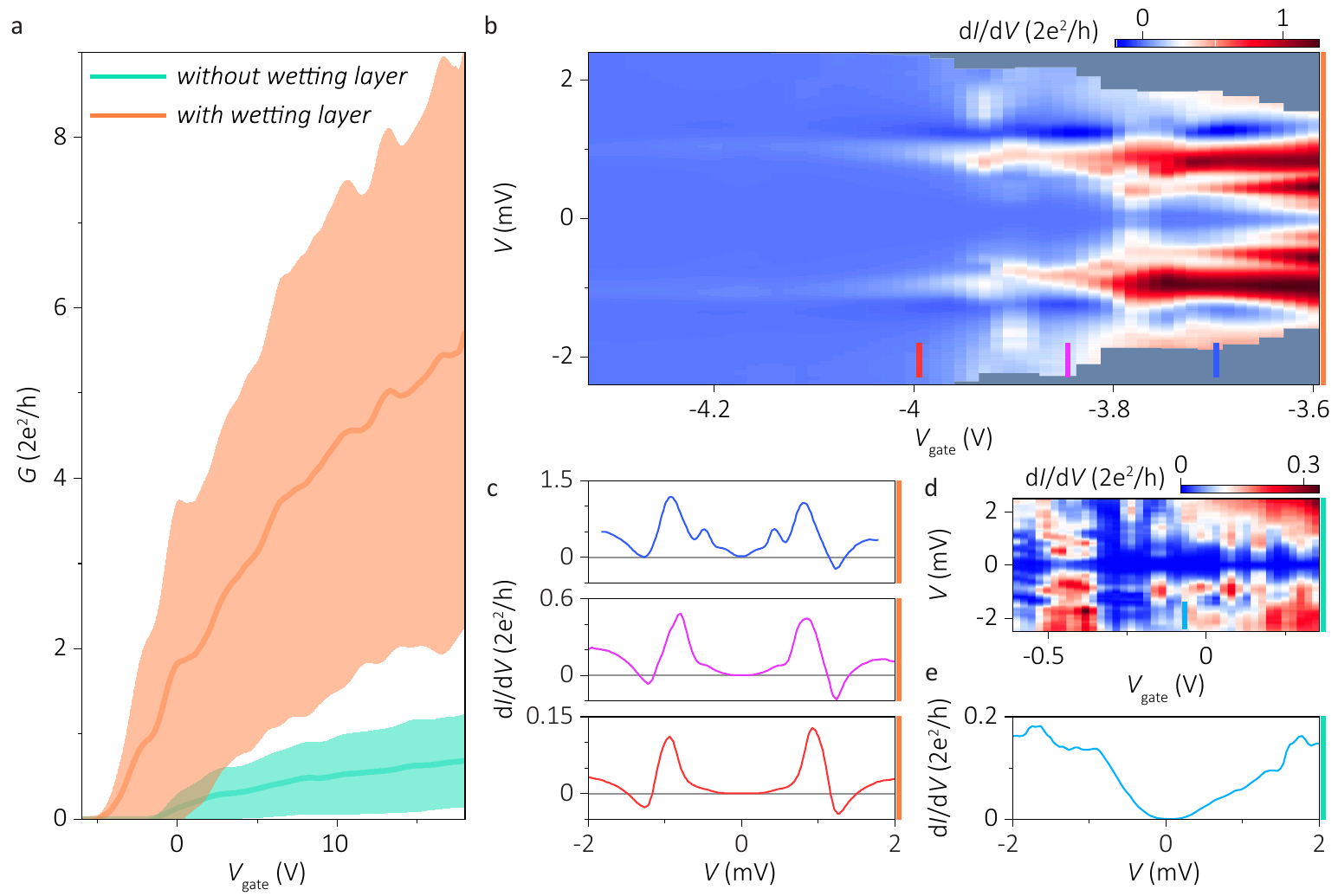}
	\caption{\footnotesize{\textbf{Effects of wetting layer on the transport and superconducting properties.} \textbf{(a)} Gate voltage dependent conductance $G$ of InSb nanowires devices with $\sim 150$ nm electrode separation realized with and without including a NbTi (5 nm) wetting layer between the nanowire and NbTiN (90 nm) electrodes. Native oxide on the nanowire surface is removed by sulfur cleaning prior to the deposition of the electrodes. Traces represent ensemble-averaged conductance over 4 (NbTi/NbTiN) and 7 (NbTiN) different devices measured at a bias voltage $V = 10$ mV, with the shades indicating the standard deviation (see the Supporting Information for the details of averaging). Inclusion of a NbTi wetting layer decreases the average contact resistance (including both contacts) from $\sim 100$ k$\Omega$ to $\sim 1.6$ k$\Omega$ (see the Supporting Information for the extraction of contact resistance). \textbf{(b), (c)} Spectroscopy of a device realized with NbTi/NbTiN electrodes. Differential conductance d$I$/d$V$ is plotted as a function of bias voltage $V$ for varying gate voltages $V_\mathrm{gate}$. d$I$/d$V$ traces in (c) are vertical line cuts from (b) at gate voltages marked with colored bars. d$I$/d$V$ is symmetric in bias with two peaks at $V \sim \pm 1$ mV seen for all gate voltages from which we find $2\Delta \sim 1$ meV. For low $V_\mathrm{gate}$ and away from quantum dot resonances subgap conductance vanishes, revealing a hard induced gap. Larger gate voltages decrease the tunnel barrier height, where increased Andreev reflection probability results in finite subgap conductance. \textbf{(d)} Spectroscopy of a device realized with NbTiN electrodes without a NbTi wetting layer. Tunneling conductance is dominated by Coulomb blockade with irregular diamonds. Induced gap cannot be clearly identified. \textbf{(e)} A vertical line cut from (d) at $V_\mathrm{gate} \sim -0.08$ V (indicated by a blue bar) with a conductance similar to the middle panel in (c). d$I$/d$V$ is not symmetric in bias and coherence peaks are not visible. All data in this figure taken at $T = 250$ mK.}}
	\label{fig3}
\end{figure}

Next, we investigate the wetting of the superconductor on the nanowire surface. Figure 3a shows the conductance averaged over different nanowire devices realized with and without a thin layer of NbTi (5 nm), a reactive metal deposited immediately before the NbTiN to ensure its wetting on the wire. Inclusion of a NbTi wetting layer substantially improves the contact resistance of the devices. Tunneling spectroscopy (Figure 3b-d) reveals the differences in superconducting properties of the devices with and without the wetting layer. Figure 3b shows an induced gap $2\Delta \sim 1$ meV for a device with NbTi wetting layer. Low gate voltages bring the device into the tunneling regime revealing a hard gap, shown in Figure 3c. In contrast, Figure 3d and e show that omitting the wetting layer results in no clearly identifiable induced gap and a tunneling conductance dominated by Coulomb blockade with irregular diamonds. Finally, to verify the importance of the wetting of the superconductor on the wire surface we realized InSb-Al nanowire devices without a Ti wetting layer. These devices also showed very high contact resistances, while inclusion of Ti wetting layer gave low contact resistances and a finite supercurrent (Figure S3), in addition to a hard gap shown in Figure 1c and d. In the Supporting Information we comprehensively discuss our observations related to the improvement due to inclusion of a wetting layer.

The devices prepared with sulfur cleaning and NbTi/NbTiN electrodes in Figure 3 did not show a supercurrent, a requirement for a nanowire-based topological quantum bit\cite{Hyart2013PRB,Plugge2016ArXiv,VijayFu2016Arxiv,KarzigHexon2016Arxiv}. We attribute the lack of a supercurrent to a residual interface barrier effective at small bias. This could be related to the \textit{ex-situ} nature of sulfur cleaning, leaving the wire surface exposed to ambient which cannot exclude adsorbents at the interface. To improve the small bias response of our devices we perform an additional \textit{in-situ} argon cleaning of sufficiently low-power to avoid a damage to the InSb nanowire surface. After including this low-power argon cleaning we find a high yield of devices showing a finite supercurrent measured at 250 mK (Figure S4). For another chip with 18 nanowire devices but measured at 50 mK, we find a clear supercurrent for all devices (Figure S5) while obtaining an induced gap $2\Delta \sim 1$ meV or larger (Figure S6 and S9).

\begin{figure}
	\vspace{-20pt}
	\includegraphics[width=0.95\columnwidth]{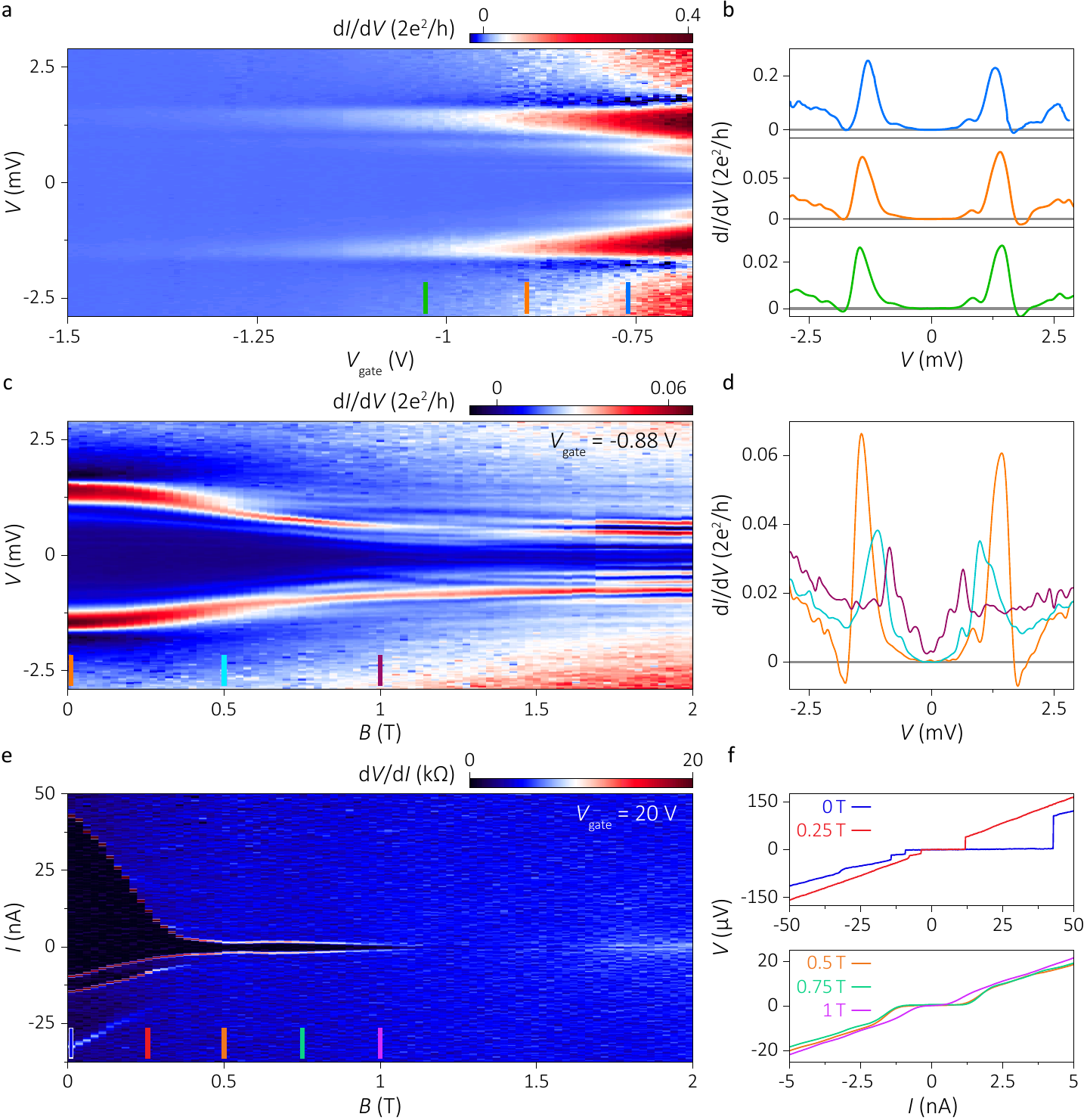}
	\caption{\footnotesize{\textbf{Tunneling spectroscopy and magnetic field response of InSb nanowire hybrid devices with engineered interface.} \textbf{(a), (b)} Spectroscopy of a device realized with NbTi/NbTiN electrodes using sulfur cleaning followed by an \textit{in-situ} low-power argon cleaning. Differential conductance d$I$/d$V$ is plotted as a function of bias voltage $V$ for varying gate voltages $V_\mathrm{gate}$. d$I$/d$V$ traces in (b) are vertical line cuts from (a) at gate voltages marked with colored bars. d$I$/d$V$ is symmetric in bias with two peaks at $V \sim \pm 1.5$ mV seen for all gate voltages from which we find $2 \Delta \sim 1.5$ meV. The induced gap is hard with vanishing subgap conductance in the tunneling regime. \textbf{(c), (d)} d$I$/d$V$ of the same device is plotted as a function of bias voltage $V$ for an increasing magnetic field $B$ along the nanowire. Gate voltage is set to $V_\mathrm{gate} = -0.88$ V, the same as in the middle panel in (b). d$I$/d$V$ traces in (d) are vertical line cuts from (c) at magnetic fields marked with colored bars. The induced gap remains hard up to $\sim 0.5$ T. Increasing fields decrease the induced gap size and increase the subgap conductance but induced superconductivity persists up to 2 T where d$I$/d$V$ shows a gap feature with suppressed conductance at small bias and symmetrically positioned coherence peaks. \textbf{(e)} Differential resistance d$V$/d$I$ of an identical device is plotted as a function of bias current $I$ for an increasing magnetic field $B$ along the nanowire. Dark regions with vanishing resistance indicate the supercurrent which remains finite up to 1 T. Gate voltage $V_\mathrm{gate} = 20$ V. \textbf{(f)} Current-voltage traces from (e) at magnetic fields marked with colored bars. We find a switching current of $\sim 40$ nA at zero magnetic field, which decreases to $\sim 10$ nA at 0.25 T, and to $\sim 0.5$ nA at 1 T. Both devices in this figure have an electrode separation of $\sim 150$ nm.} Data taken at $T = 50$ mK.}
	\label{fig4}
\end{figure}

Finally we study the magnetic field response of the optimized hybrid devices combining sulfur cleaning followed by an \textit{in-situ} low-power argon cleaning, and NbTi/NbTiN superconducting electrodes. Figure 4a and b show the differential conductance for varying gate voltages at zero magnetic field measured at 50 mK (details in Figure S6). We find a hard gap $2\Delta \sim 1.5$ meV which confirms the noninvasiveness of our low-power cleaning. The extracted conductance suppression at small bias compared to the above-gap conductance at large bias is $\sim 100$ (Figure S7). Next, we choose a gate voltage where the device is in the tunneling regime (orange trace in Figure 4b) and perform spectroscopy for increasing magnetic fields along the wire axis, shown in Figure 4c. In Figure 4d we plot the conductance traces taken at different magnetic fields showing an induced gap which remains hard up to $\sim 0.5$ T (see Figure S8 for a logarithmic plot). Increasing fields decrease the induced gap size and increase the subgap conductance, but a gap feature can be identified up to 2 T revealing the large critical field of NbTiN. Figure 4e and f show the critical current of another device as a function of magnetic field, measured at a large gate voltage when the nanowire is highly conducting (details in Figure S9). We find a critical current of $\sim 40$ nA at zero magnetic field which remains finite up to greater than 1 T. The nonmonotonous magnetic-field evolution of the critical current can be accounted for using a model which includes Zeeman effect, spin-orbit coupling, and a realistic nanowire geometry in the few-channel, quasi-ballistic regime -- the transport regime of our devices\cite{Zuo2017}.

In conclusion, we have developed a method of obtaining a hard induced gap and supercurrent in InSb nanowires in the presence of magnetic fields ($\sim 0.5$ Tesla) by combining a noninvasive nanowire surface cleaning together with a wetting layer between the nanowire and the NbTiN superconductor. Our results provide a guideline for inducing superconductivity in semiconductor nanowires, two dimensional electron gases and topological insulators, and hold relevance for topological superconductivity in various material systems.

\begin{acknowledgement}

We thank S. Goswami and J. Shen for stimulating discussions and critical reading of the manuscript, and D.B. Szombati for assistance in device fabrication. This work has been supported by the Netherlands Organisation for Scientific Research (NWO), Foundation for Fundamental Research on Matter (FOM), European Research Council (ERC), Office of Naval Research (ONR N00014-16-1-2270), and Microsoft Corporation Station Q.

\end{acknowledgement}

%

\bibliography{hardgap}

\clearpage
\newpage
\renewcommand\thefigure{S\arabic{figure}}
\renewcommand{\bibnumfmt}[1]{(S#1)}
\setlength\parindent{0pt}
\linespread{1.2}

\begin{center}
\textbf{\large Supporting Information: Hard superconducting gap in InSb nanowires}
\end{center}

\vspace{0cm}

\begin{center}
	\normalsize{
		\"Onder~G\"ul$^{1,2,*}$, Hao~Zhang$^{1,2}$, Folkert~K.~de~Vries$^{1,2}$, Jasper~van~Veen$^{1,2}$, Kun~Zuo$^{1,2}$, Vincent~Mourik$^{1,2}$, Sonia~Conesa-Boj$^{1,2}$, Micha\l{}~P.~Nowak$^{1,2,3}$, David~J.~van~Woerkom$^{1,2}$, Marina~Quintero-P\'erez$^{1,4}$, Maja~C.~Cassidy$^{1,2}$, Attila~Geresdi$^{1,2}$, Sebastian~Koelling$^{5}$, Diana~Car$^{1,2,5}$, S\'ebastien~R.~Plissard$^{2,5,6}$, Erik~P.A.M.~Bakkers$^{1,2,5}$, Leo~P.~Kouwenhoven$^{1,2,7,\dagger}$
			   }
\end{center}

\small
\begin{center}
	\emph{$^\mathit{1}$QuTech, Delft University of Technology, 2600 GA Delft, The Netherlands}
	
	\emph{$^\mathit{2}$Kavli Institute of Nanoscience, Delft University of Technology, 2600 GA Delft, The Netherlands}
	
	\emph{$^\mathit{3}$Faculty of Physics and Applied Computer Science, AGH University of Science and Technology, al. A. Mickiewicza 30, 30-059 Krak\'ow, Poland}
	
	\emph{$^\mathit{4}$Netherlands Organisation for Applied Scientific Research (TNO), 2600 AD Delft, The Netherlands}
	
	\emph{$^\mathit{5}$Department of Applied Physics, Eindhoven University of Technology, 5600 MB Eindhoven, The Netherlands}
	
	\emph{$^\mathit{6}$CNRS-Laboratoire d'Analyse et d'Architecture des Syst\`emes (LAAS), Universit\'e de Toulouse, 7 avenue du colonel Roche, F-31400 Toulouse, France}
		
	\emph{$^\mathit{7}$Microsoft Station Q Delft, 2600 GA Delft, The Netherlands}
\end{center}

\begin{center}
	$^*$E-mail: gulonder@gmail.com
	$^\dagger$E-mail: L.P.Kouwenhoven@TUDelft.nl
\end{center}

\vspace{1cm}

\textbf{Author contributions.} \"OG, FKdV, KZ, and VM developed the noninvasive surface cleaning and the inclusion of wetting layer. \"OG, HZ, and JvV optimized the noninvasive surface cleaning. SCB did the TEM analysis. MPN did the theoretical analysis. DJvW contributed to device fabrication. MQP and MCC optimized the NbTiN films. AG contributed to the data analysis. SK prepared the lamellae for TEM analysis. DC, SRP, and EPAMB grew the InSb nanowires. \"OG wrote the manuscript with contributions from all authors. LPK supervised the project.

\clearpage
\newpage

\normalsize
\begin{center}\large{
	\textbf{List of supporting text and figures}}
\end{center}

\vspace{0cm}

\begin{raggedright}
\begin{itemize}

\item[] \textbf{Nanowire growth and device fabrication}

\item[] \textbf{Fabrication details of the long-channel devices in Figure 2}

\item[] \textbf{Measurement setup}

\item[] \textbf{Details of ensemble averaging}

\item[] \textbf{Extraction of contact resistance}

\item[] \textbf{Discussion of the wetting layer}

\item[] \textbf{Figure S1:} Magnetic field response of the induced gap in InSb nanowire hybrid device with Ti/Al electrodes

\item[] \textbf{Figure S2:} Cross-sectional transmission electron micrographs of the nanowire surface cleaned using different methods

\item[] \textbf{Figure S3:} Additional transport properties of InSb nanowire hybrid devices with Ti/Al electrodes

\item[] \textbf{Figure S4:} Supercurrent in InSb nanowire hybrid devices with NbTi/NbTiN electrodes

\item[] \textbf{Figure S5:} Supercurrent in InSb nanowire hybrid devices with NbTi/NbTiN electrodes at T = 50 mK

\item[] \textbf{Figure S6:} Additional transport properties of InSb nanowire hybrid device with engineered interface (device A)

\item[] \textbf{Figure S7:} Figure 4b replotted in logarithmic conductance scale

\item[] \textbf{Figure S8:} Figure 4d replotted in logarithmic conductance scale

\item[] \textbf{Figure S9:} Additional transport properties of InSb nanowire hybrid device with engineered interface (device B)

\end{itemize}
\end{raggedright}

\clearpage
\newpage

\textbf{Nanowire growth and device fabrication.} InSb nanowires have been grown by Au-catalyzed Vapor-Liquid-Solid mechanism in a Metal Organic Vapor Phase Epitaxy reactor. The InSb nanowire crystal direction is [111] zinc blende, free of stacking faults and dislocations\cite{s1}. Nanowires are deposited one-by-one using a micro-manipulator\cite{s2} on a p-Si++ substrate covered with 285 nm thick SiO$_2$ serving as a dielectric for back gate. Superconductor deposition process starts with resist development followed by oxygen plasma cleaning. For sulfur cleaning, the chip is immersed in a Sulfur-rich ammonium sulfide solution diluted by water (with a ratio of 1:200) at $60^{\circ}$C for half an hour\cite{s3}. At all stages care is taken to expose the solution to air as little as possible. Ti/Al contacts are e-beam evaporated at a base pressure $< 10^{-7}$ mbar. In-situ argon plasma cleaning and NbTiN deposition is performed in an AJA International ATC 1800 sputtering system with a base pressure $\sim 10^{-9}$ Torr. For devices without sulfur cleaning, argon cleaning is performed using an argon plasma typically at a pressure of 3 mTorr and a power of 100 Watts applied for 150 seconds, but different plasma parameters removing a similar thickness of InSb from the nanowire surface ($> 15$ nm) gave similar transport properties. For devices with sulfur cleaning we used a much milder argon plasma at a pressure of 10 mTorr and a power of 25 Watts applied for $\sim 5$ seconds. For NbTiN deposition a Nb$_{0.7}$Ti$_{0.3}$ wt. \% target with a diameter of 3 inches is used. Reactive sputtering resulting in (NbTi) NbTiN films was performed in an Ar/N process gas with (0) 8.3 at. \% nitrogen content at a pressure of 2.5 mTorr using a dc magnetron sputter source at a power of 250 Watts. An independent characterization of the NbTiN films gave a critical temperature of 13.5 K for 90 nm thick films with a resistivity of 114 $\mu \Omega \cdot$cm and a compressive stress on Si substrate.

\bigskip

\textbf{Fabrication details of the long-channel devices in Figure 2.} For the InSb nanowire devices with sulfur-cleaned channels, the cleaning of the channel is performed after a complete fabrication of the electrodes contacting the nanowire. For the devices whose transport data is presented, sulfur cleaning is applied to the entire channel, while the inset shows a partially cleaned channel to illustrate the mild etching of the wire. For the nanowire devices with argon-cleaned channels, the cleaning of the channel is performed before the fabrication of the contact electrodes. However, we obtained a similar result when argon cleaning was applied after fabricating the contacts. For all long-channel devices we used argon cleaning to remove the native oxide on the nanowire surface prior to contact deposition.

\bigskip

\textbf{Measurement setup.} All the data in this study is measured using RC, copper powder, and $\pi$ filters thermalized at different temperatures. Differential conductance measurements are performed using standard ac lock-in techniques. Nanowire devices are kept in vacuum during low temperature measurements.

\bigskip

\textbf{Details of ensemble averaging.} Conductance is averaged over different nanowire devices for each value of gate voltage. Devices within an ensemble are fabricated simultaneously on a single substrate, have identical geometries, and are measured during the same cool down.

\bigskip

\textbf{Extraction of contact resistance.} We extract contact resistances by fitting the conductance measured as a function of gate voltage using the method described in Ref. S4. Here we leave the product of capacitance and mobility as a free fit parameter which is not taken into consideration.

\bigskip

\textbf{Discussion of the wetting layer.} Throughout our study we have tried various etching techniques (HF, lactic acid, sulfur solution, He ion beam, Ar plasma) in combination with different contact materials (Ti, Al, V, Cr - evaporation; Al, NbTi, NbTiN, MoRe - sputtering). Our observations rule out a work function ($W$) related explanation for the improvement due to inclusion of a wetting layer: Ti ($W = 4.33$ eV)\cite{s5} and Cr ($W = 4.5$ eV)\cite{s5} gives significantly lower contact resistances on our wires than Al ($W = 4.06$ -- $4.26$ eV)\cite{s5} -- known for its low work function. (See Ref. S6 for Cr/Au contacts on our wires.) Consistent observation of the improvement due to wetting layer both for evaporation (Al) and for reactive sputtering (NbTiN) suggests that the mechanism of improvement is independent of the details of the deposition environment, such as nitrogen plasma. We further note that sputtered Al contacts also result in low contact transparency with large contact resistances (no difference observed between evaporated and sputtered Al contacts without a wetting layer). Finally, a Ti wetting layer provides improvement for Al contacts on InAs nanowires as well\cite{s7} -- consistent with our observations. However, it is possible to realize transparent contacts of Al on InAs when the Al-InAs interface is homogeneous (epitaxial)\cite{s7}. This suggests that the nucleation of the Al film deposited on the wire is a key factor which determines the transparency.

\clearpage
\newpage

\begin{figure}[H]
	\centering
	\includegraphics[width=1\columnwidth]{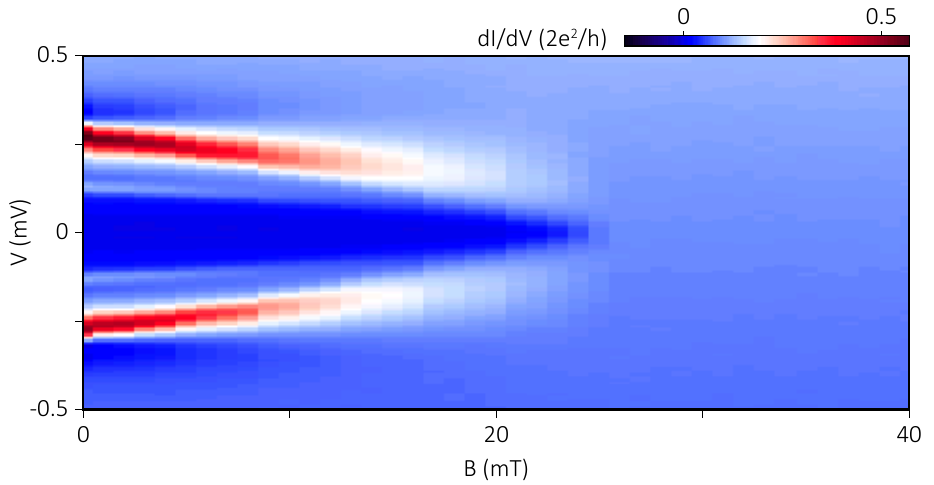}
\end{figure}

Figure S1: \textbf{Magnetic field response of the induced gap in InSb nanowire hybrid device with Ti/Al electrodes.}
Differential conductance d$I$/d$V$ is plotted as a function of bias voltage $V$ for increasing magnetic field $B$ along the nanowire axis. The nanowire device is in the tunneling regime with d$I$/d$V$ $\ll 2e^2/h$ for above-gap bias ($V > 2\Delta$). At zero magnetic field d$I$/d$V$ shows two conductance peaks at $V \sim \pm 0.3$ mV symmetric around zero bias giving $2\Delta \sim 0.3$ meV. Increasing the magnetic field decreases the size of the superconducting gap $\Delta$ which completely vanishes at $\sim 25$ mT. The device shows no superconductivity at larger magnetic fields. Device realized by sulfur cleaning. Ti/Al electrodes have a thickness of 5/130 nm and a separation of $\sim 150$ nm on the nanowire. $T = 50$ mK.

\clearpage

\begin{figure}
	\centering
	\includegraphics[width=1\textwidth]{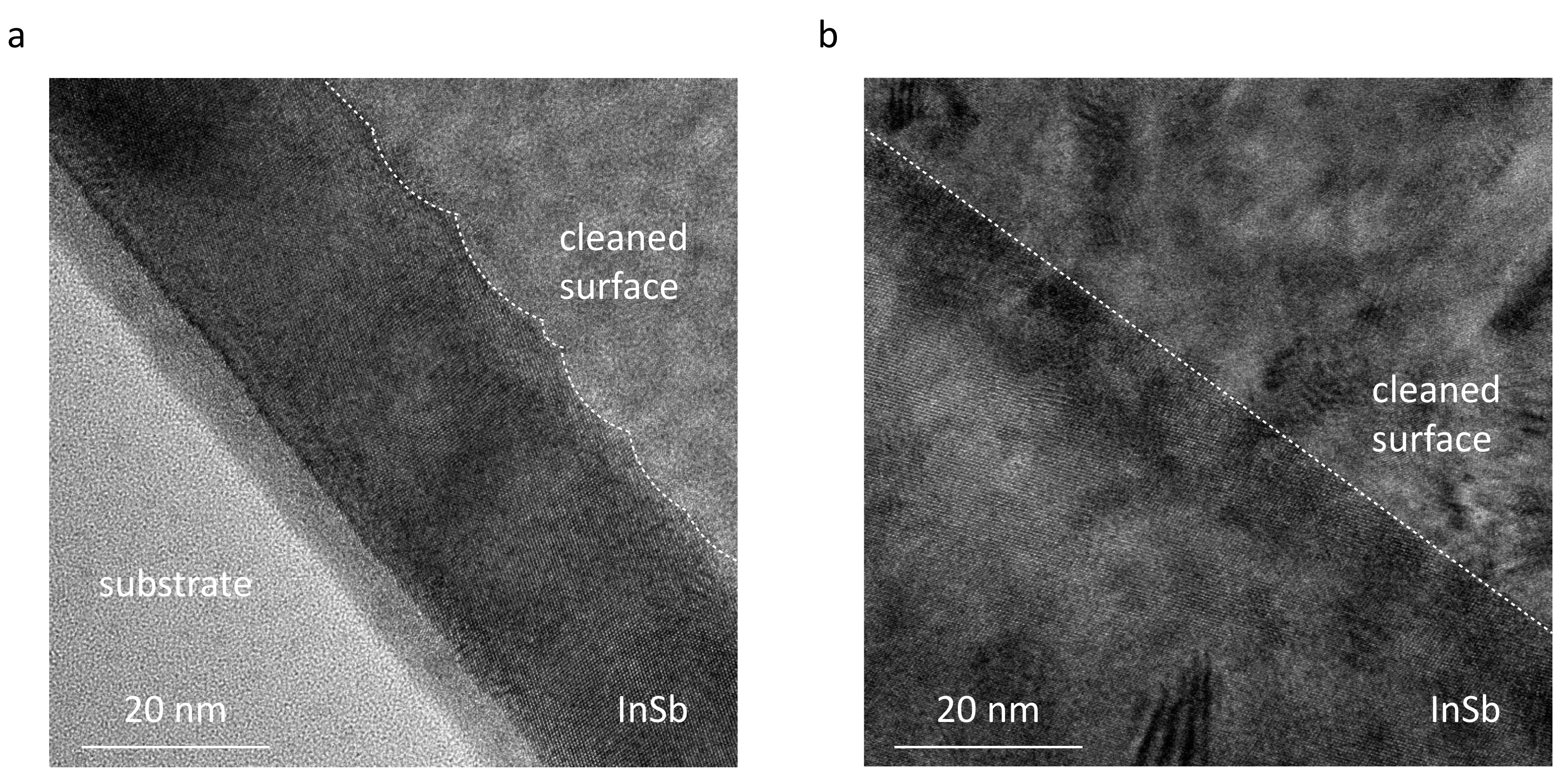}
\end{figure}

Figure S2: \textbf{Cross-sectional transmission electron micrographs of the nanowire surface cleaned using different methods.}
The cuts were performed along the nanowire axis. \textbf{(a)} InSb nanowire surface after argon plasma cleaning. Argon cleaning leaves a rough nanowire surface. Nanowire appears thinner due to substantial etching. \textbf{(b)} InSb nanowire surface after sulfur cleaning followed by a low-power argon cleaning (see main text for details). Sulfur cleaning leaves a smoother nanowire surface.

\clearpage

\begin{figure}[H]
	\centering
	\includegraphics[width=0.8\textwidth]{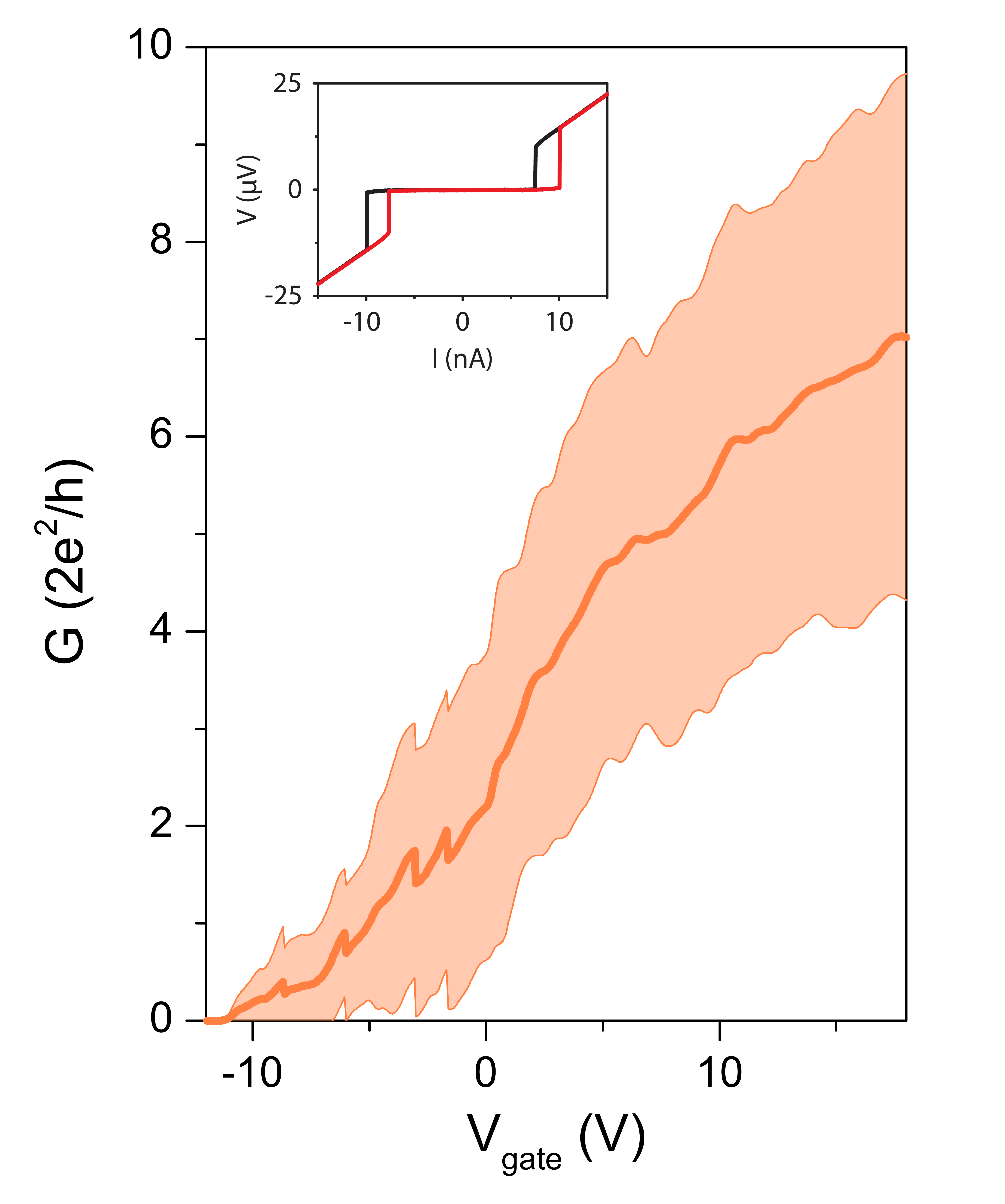}
\end{figure}

Figure S3: \textbf{Additional transport properties of InSb nanowire hybrid devices with Ti/Al electrodes.}
Conductance $G$ of InSb nanowire devices is plotted as a function of gate voltage $V_\mathrm{gate}$. The trace represents ensemble-averaged conductance over 6 different devices on a single chip and the shade indicates the standard deviation. Taken at a bias voltage $V = 10$ mV. Average contact resistance (including both contacts) is $\sim 1$ k$\Omega$. Inset shows voltage drop $V$ as a function of bias current $I$ for one of the 6 devices. We find a clear supercurrent up to $\sim 10$ nA. Red trace shows the current-voltage response when the bias current is swept in positive direction, black trace the negative direction. Gate voltage is set to $V_\mathrm{gate} = 9$ V. Ti/Al electrodes have a thickness of 5/130 nm and a separation of $\sim 150$ nm on the nanowire. All data in this figure taken at $T = 250$ mK.

\clearpage

\begin{figure}[H]
	\centering
	\includegraphics[width=0.9\textwidth]{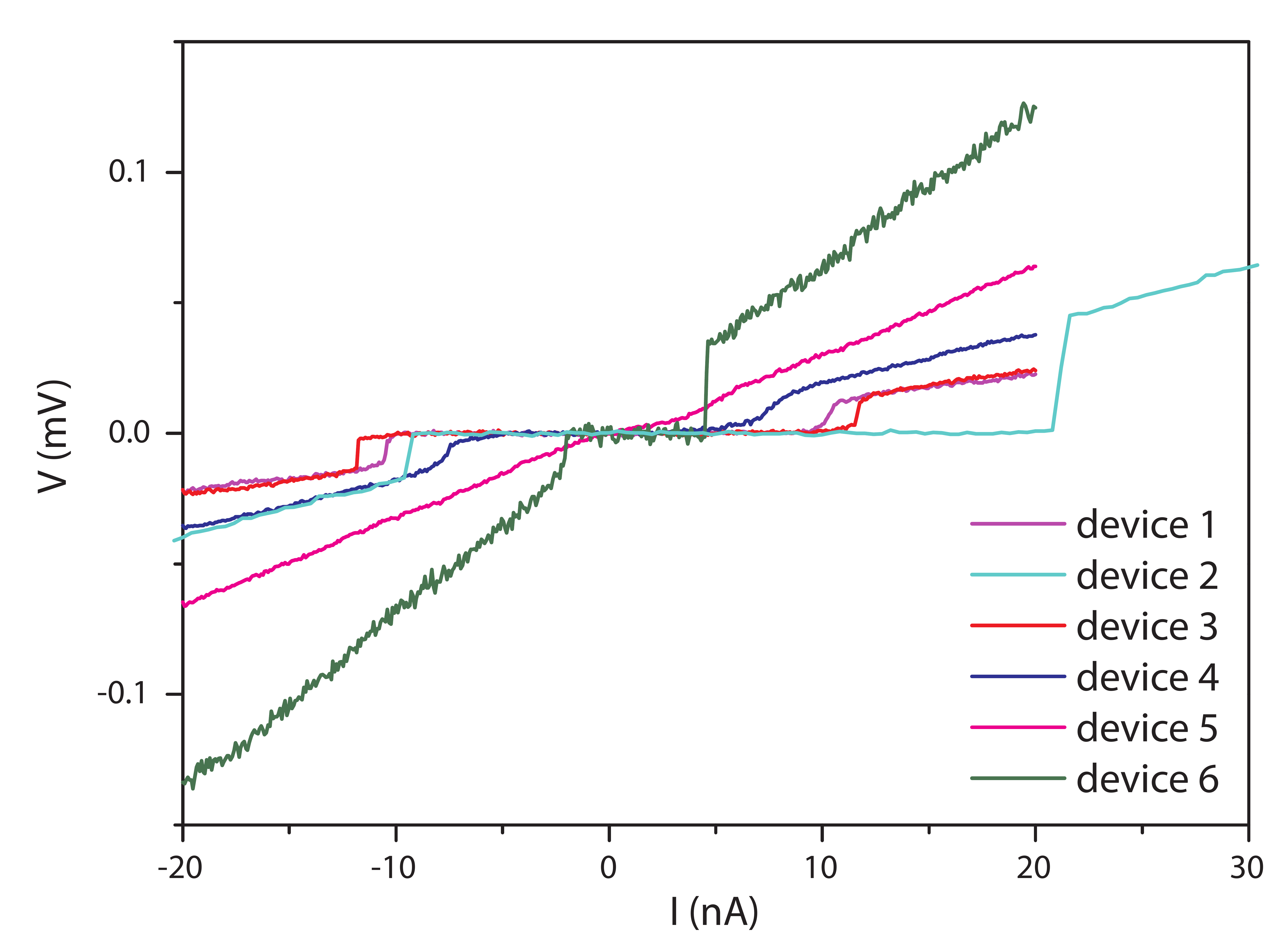}
\end{figure}

Figure S4: \textbf{Supercurrent in InSb nanowire hybrid devices with NbTi/NbTiN electrodes.}
Devices realized using sulfur cleaning followed by an \textit{in-situ} low-power argon cleaning. $T = 250$ mK. Voltage drop $V$ is plotted as a function of bias current $I$ for different devices on a single chip with 8 devices in total. An \textit{in-situ} low-power argon cleaning improves the small bias response allowing to resolve a supercurrent for a high yield of devices. Gate voltage is set to $V_\mathrm{gate} = 18$ V. NbTi/NbTiN electrodes have a thickness of 5/90 nm and a separation of $\sim 150$ nm on the nanowire.

\clearpage

\begin{figure}[H]
	\centering
	\includegraphics[width=1\textwidth]{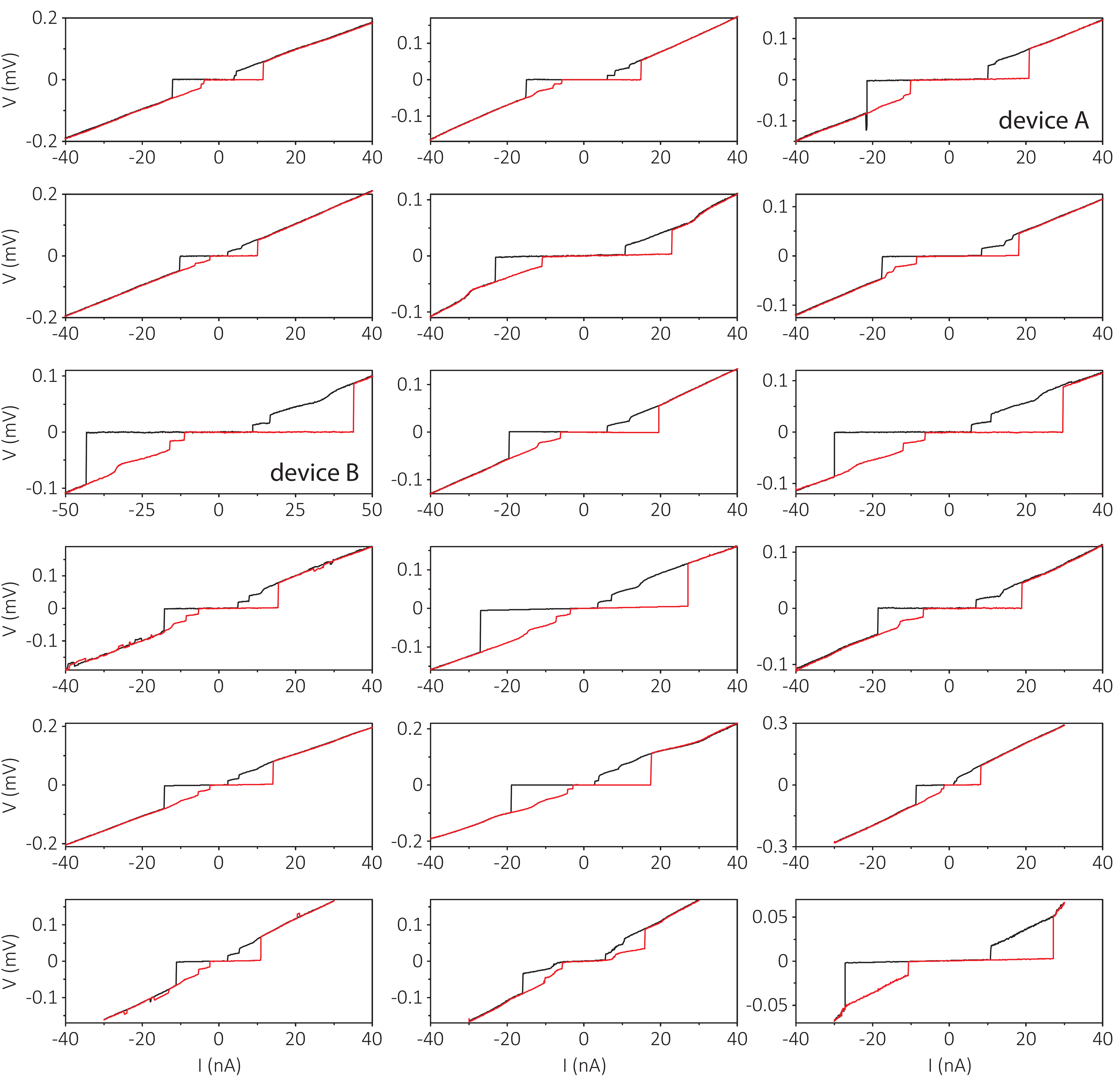}
\end{figure}

Figure S5: \textbf{Supercurrent in InSb nanowire hybrid devices with NbTi/NbTiN electrodes at T = 50 mK.}
Devices realized using sulfur cleaning followed by an \textit{in-situ} low-power argon cleaning. Voltage drop $V$ is plotted as a function of bias current $I$ for all devices on a single chip. We find a clear supercurrent in every device. Red traces show the current-voltage response when the bias current is swept in positive direction, black traces the negative direction. We relate the origin of the hysteresis to electron heating\cite{s8}. Gate voltage is set to $V_\mathrm{gate} = 20$ V. NbTi/NbTiN electrodes have a thickness of 5/90 nm and a separation of $\sim 150$ nm on the nanowire. Data in Figure 4a-d is taken from Device A, data in Figure 4e,f taken from Device B.

\clearpage

\begin{figure}[H]
	\centering
	\includegraphics[width=1.0\columnwidth]{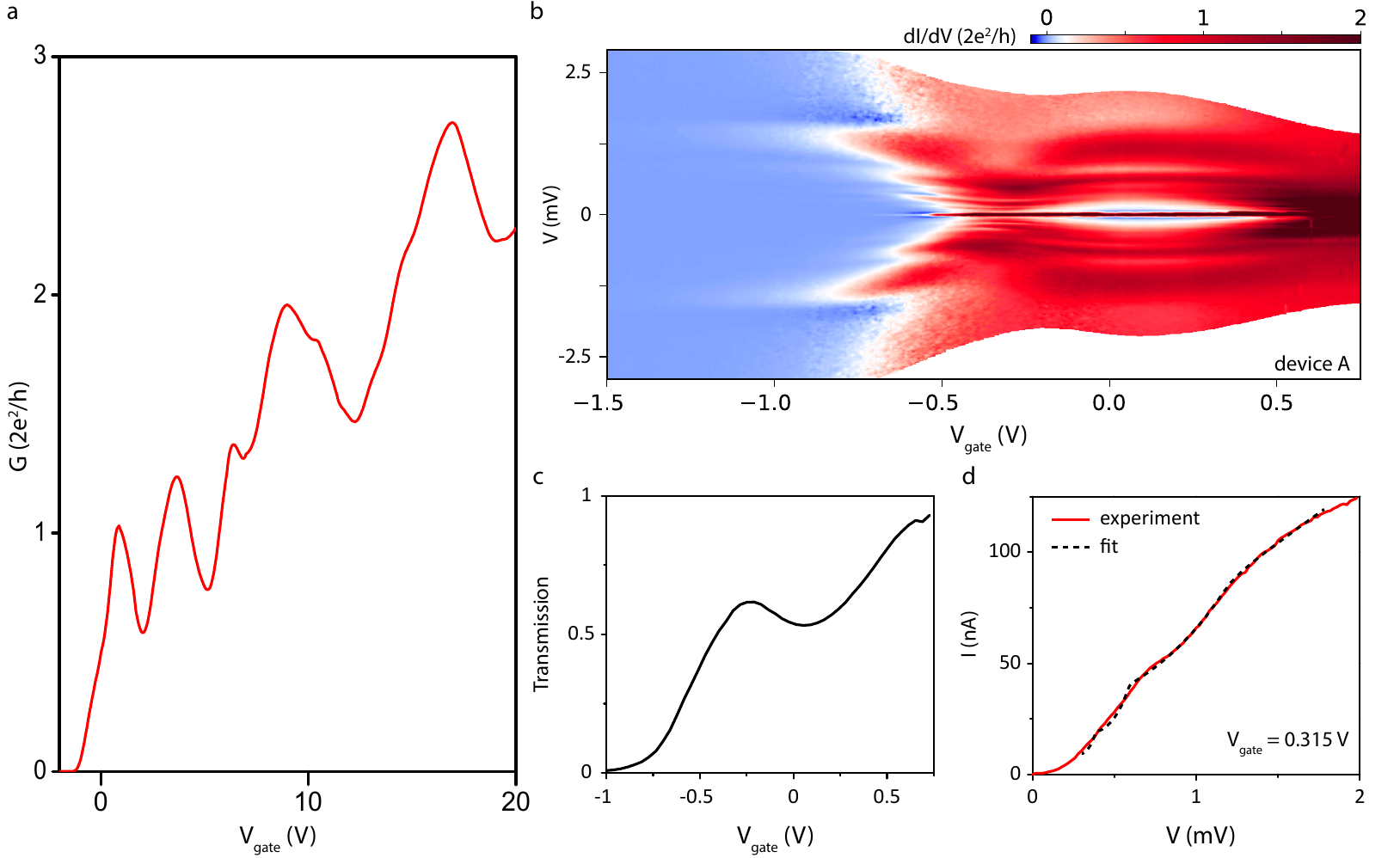}
\end{figure}

Figure S6: \textbf{Additional transport properties of InSb nanowire hybrid device with engineered interface (device A).}
All data in this figure is taken from device A, the device in Figure 4a-d. \textbf{(a)} Conductance $G$ of InSb nanowire device is plotted as a function of gate voltage $V_\mathrm{gate}$, taken at a bias voltage $V = 10$ mV. Extracted contact resistance (including both contacts) $\sim 1.6$ k$\Omega$. \textbf{(b)} Differential conductance d$I$/d$V$ is plotted as a function of bias voltage $V$ for a large gate voltage $V_\mathrm{gate}$ range. \textbf{(c)} Transmission $T$ extracted from (b) by fitting the measured current $I$ using Averin-Bardas model\cite{s9,s10}. Best fit is obtained for a single channel transport through the nanowire. Transmission reaches $\sim 0.9$ giving a lower bound on contact transparency assuming a single channel transport. We note that our model does not account for the observed finite contact transparencies and conductance resonances, decreasing the certainty of our estimate. \textbf{(d)} The measured current $I$ and the corresponding fit is plotted for $V_\mathrm{gate} = 0.315$ V.

\clearpage

\begin{figure}[H]
	\centering
	\includegraphics[width=1\columnwidth]{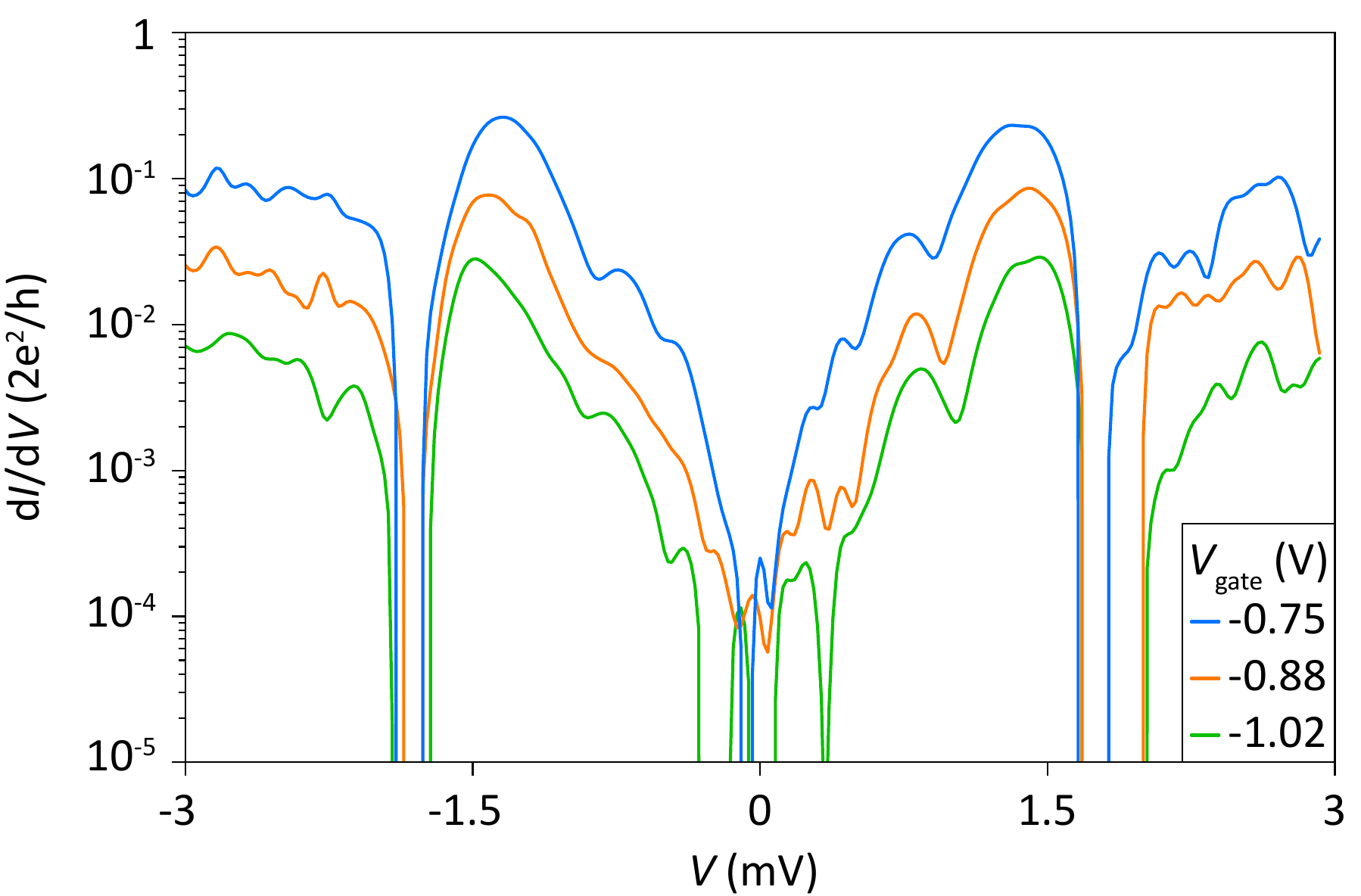}
\end{figure}

Figure S7: \textbf{Figure 4b replotted in logarithmic conductance scale.} Differential conductance d$I$/d$V$ is plotted as a function of bias voltage $V$ for varying gate voltages $V_\mathrm{gate}$. The extracted conductance suppression at small bias compared to the above-gap conductance at large bias is $\sim 100$.

\clearpage

\begin{figure}[H]
	\centering
	\includegraphics[width=1\columnwidth]{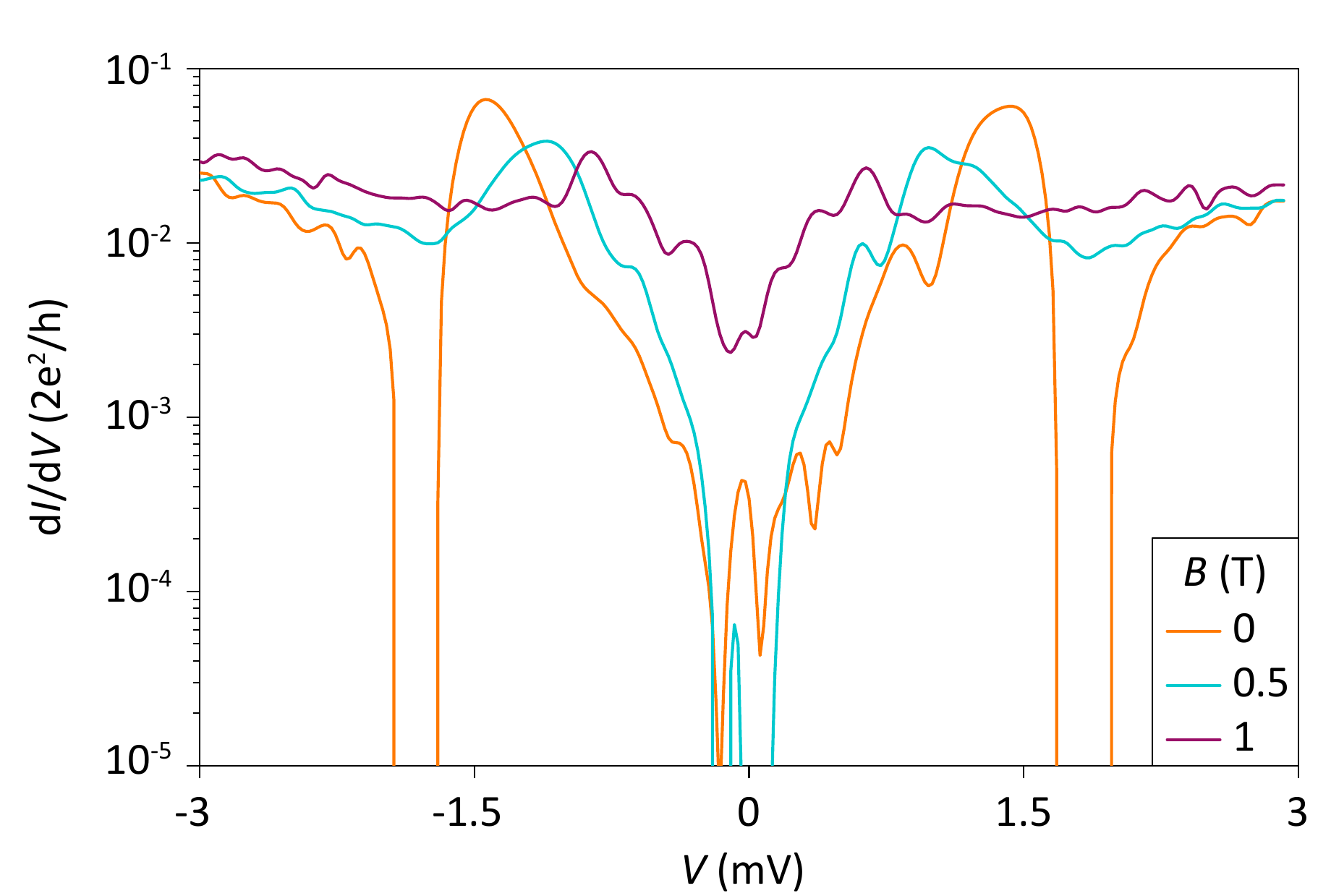}
\end{figure}

Figure S8: \textbf{Figure 4d replotted in logarithmic conductance scale.} Differential conductance d$I$/d$V$ is plotted as a function of bias voltage $V$ for varying magnetic fields $B$ along the wire axis. The induced gap remains hard up to $\sim 0.5$ T.

\clearpage

\begin{figure}[H]
	\centering
	\includegraphics[width=1\columnwidth]{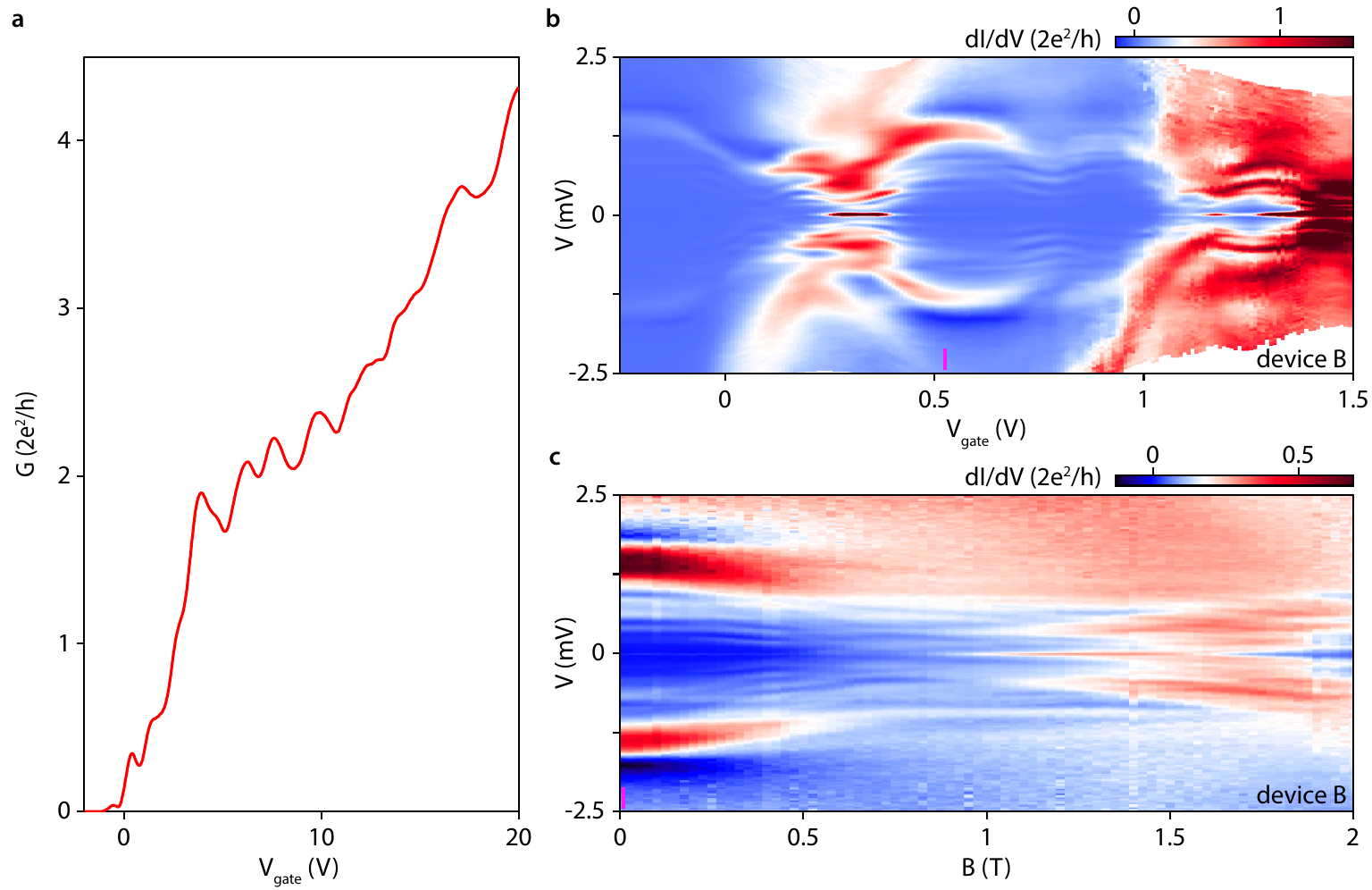}
\end{figure}

Figure S9: \textbf{Additional transport properties of InSb nanowire hybrid device with engineered interface (device B).}
All data in this figure is taken from device B, the device in Figure 4e,f. \textbf{(a)} Conductance $G$ of InSb nanowire device is plotted as a function of gate voltage $V_\mathrm{gate}$, taken at a bias voltage $V = 10$ mV. Extracted contact resistance (including both contacts) $\sim 1.2$ k$\Omega$. \textbf{(b)} Differential conductance d$I$/d$V$ is plotted as a function of bias voltage $V$ for varying gate voltages $V_\mathrm{gate}$. Differential conductance shows quantum dot features similar to those previously reported for InSb nanowires\cite{s11}. Further, we find subgap conductance peaks running through consecutive Coulomb valleys which we attribute to Andreev bound states in the wire section underneath the superconducting electrodes\cite{s12}. For our device geometry with a back gate controlling both the conductance in the wire section between the electrodes as well as the occupation in the wire section underneath the electrodes, it is not possible to tune the device away from the subgap states while maintaining the tunnelling regime necessary for spectroscopy. \textbf{(c)} Differential conductance d$I$/d$V$ is plotted as a function of bias voltage $V$ for an increasing magnetic field $B$ along the nanowire. $V_\mathrm{gate} \sim 0.5$ V indicated with a pink bar in (b). Increasing magnetic fields bring the subgap states to lower energies resulting in a finite subgap conductance for $B > 0.3$ Tesla, similar to a previous report\cite{s12} but at a relatively lower magnetic field.

\end{document}